\input{aipcheck}
\documentclass[
    ,final            
  ]
  {aipproc}

\layoutstyle{8x11single}

\begin{document}
\title{The Theory of R-parity, Unification and SUSY at the LHC}
%
\author{Pavel Fileviez P\'erez}
{ address={ Center for Cosmology and Particle Physics (CCPP) 
\\ 
New York University, 4 Washington Place, NY 10003, USA}}
\begin{abstract}
The simplest gauge theories for the conservation of R-parity in supersymmetry are discussed.
We show how the minimal theory based on the B-L gauge symmetry predicts that R-parity must be 
spontaneously broken at the TeV scale. The most striking signals of these theories at the Large Hadron Collider 
are discussed. We present a realistic theory where the local baryon and lepton numbers are spontaneously broken at the supersymmetry breaking scale. 
The possibility to understand the conservation of R-parity in grand unified theories defined in four dimensions is mentioned.
\end{abstract}
\maketitle
\section{Introduction}
The idea of supersymmetry has been considered for more than thirty five years as one of the most interesting ideas with application to particle physics. 
The minimal supersymmetric standard model (MSSM) could be tested soon at the Large Hadron Collider. 
However, it is very difficult to test its predictions because we know nothing about the supersymmetric spectrum, and about 
the presence of the lepton (L) and baryon (B) number violating interactions, which can modify most of the MSSM predictions for collider 
physics and cosmology. In this short review I will discuss the new results presented in 
Refs.~\cite{FileviezPerez:2008sx,Barger:2008wn,FileviezPerez:2012mj,Barger:2010iv,FileviezPerez:2010ek,
Perez:2011zx,FileviezPerez:2011kd,Feldman:2011ms,FileviezPerez:2011pt,FileviezPerez:2012iw,Arnold:2012fm}.

It is well-known that the MSSM superpotential is given by
\begin{equation}
{\cal W}_{MSSM}= Y_u \ \hat{Q} \ \hat{H_u} \ \hat{u}^c \ + \  Y_d \ \hat{Q} \ \hat{H_d} \ \hat{d}^c \ + \  Y_e \ \hat{L} \ \hat{H_d} \ \hat{e}^c \ + \  \mu \  \hat{H_u} \ \hat{H}_d 
\ + \  W_{BL} \ + \  W_{BL}^{(5)}, 
\end{equation} 
with 
\begin{equation}
{\cal W}_{BL}= \epsilon \  \hat{L} \  \hat{H}_u \ + \  \lambda \ \hat{L} \ \hat{L} \ \hat{e}^c \ + \ \lambda^{'} \ \hat{Q} \ \hat{L} \ \hat{d}^c \ + \  \lambda^{''} \ \hat{u}^c \  \hat{d}^c \  \hat{d}^c, 
\end{equation}
and
\begin{equation}
{\cal W}_{BL}^{(5)} = \frac{c_L}{\Lambda} \  \hat{Q} \ \hat{Q} \ \hat{Q} \ \hat{L} \ + \  \frac{c_R}{\Lambda} \  \hat{u}^c \ \hat{d}^c \ \hat{u}^c \ \hat{e}^c,
\end{equation}
where the first three terms in ${\cal W}_{BL}$ break L, the last term violates B, and the interactions in ${\cal W}_{BL}^{(5)}$ break both symmetries.
There are several phenomenological constraints on the B and L violating interactions, but the most important one is coming from 
proton decay. For example, if we assume that the couplings $\lambda^{'}$ and $\lambda^{''}$ are of order one and the squark 
mass around TeV, the lifetime of the proton is about $\tau_p^{(4)} \sim 10^{-20}$ years. Therefore, in order to satisfy the experimental 
bounds on the proton decay lifetime, $\tau_p > 10^{32-34}$ years, it is important to understand the possible absence of these interactions. The operators in ${\cal W}_{BL}^{(5)}$, Eq.(3), mediate the so-called dimension five contributions to proton decay~\cite{Nath} and the scale $\Lambda$ should be larger than $10^{17}$ GeV in order to satisfy the experimental bounds. Therefore, one has to assume a desert between the TeV scale and grand unified scale.

In order to avoid the interactions in ${\cal W}_{BL}$, Eq.(2), it is often assumed the discrete symmetry, $R=(-1)^{2 S} M$, where $S$ is the spin 
and
\begin{equation}
M=(-1)^{3(B-L)},
\end{equation} 
is called matter parity. Notice that in this case the interactions in ${\cal W}_{BL}^{(5)}$ are allowed and still the scale $\Lambda$ has to be very large.
Unfortunately, in most of the studies the conservation of this symmetry is enforced or the explicit breaking is considered~\cite{Barbier:2004ez}. 
Only a few groups have studied the dynamical origin of these interactions, see for example Refs.~\cite{Aulakh:1982yn,Masiero:1990uj,Hayashi:1984rd,Mohapatra:1986aw,Krauss:1988zc,Font:1989ai,Martin:1992mq,Aulakh:1999cd,Ambroso:2009jd,FileviezPerez:2008sx,Barger:2008wn,FileviezPerez:2012mj,Barger:2010iv,FileviezPerez:2010ek,Perez:2011zx,FileviezPerez:2011kd,Feldman:2011ms,FileviezPerez:2011pt,Arnold:2012fm}. The main goal of this review is to discuss the different theories where we can understand the conservation or violation of matter parity in supersymmetry. 

In order to understand the origin of the B and L violating interactions in the MSSM we can consider a theory defined at the TeV scale or we can discuss this issue in the context of a grand unified theory in four dimensions. Before we study any particular theory we have to notice the relation between matter parity and B-L. It is obvious that if we consider a theory where B-L is part of the gauge symmetry, at the B-L scale matter parity is conserved, but when the gauge symmetry is broken one has (non) conservation of $M$ if the field responsible for symmetry breaking has an (odd) even number 
of B-L.Therefore, we can say that the theories based on the B-L gauge symmetry are the simplest frameworks where we can investigate this issue.
\subsection{I. The Minimal Theory for R-parity Violation}
Let us consider the simplest supersymmetric theory based on the B-L gauge symmetry~\cite{Barger:2008wn}
\begin{displaymath}
G_{B-L}=SU(3)_C \bigotimes SU(2)_L \bigotimes U(1)_Y \bigotimes U(1)_{B-L}
\end{displaymath} 
where we have the MSSM superfields, $\hat{Q}$, $\hat{u}^c$, $\hat{d}^c$, $\hat{L}$, $\hat{e}^c$, $\hat{H}_u$, $\hat{H}_d$, 
and the right-handed neutrinos needed to cancel the B-L anomalies, $\hat{\nu}^c$. In this context the superpotential has a simple 
form   
\begin{equation}
{\cal W}_{B-L}=Y_u \ \hat{Q} \ \hat{H}_u \  \hat{u}^c \ + \  Y_d \  \hat{Q} \  \hat{H}_d \  \hat{d}^c 
\ + \  Y_e \ \hat{L} \ \hat{H}_d \ \hat{e}^c \ + \  Y_\nu \ \hat{L} \ \hat{H}_u \  \hat{\nu}^c  \ + \  \mu \  \hat{H_u} \ \hat{H}_d \ + \ W_{B-L}^{(5)}
\end{equation}
with
\begin{equation}
{\cal W}_{B-L}^{(5)} = {\cal W}_{BL}^{(5)} \ + \  \frac{\lambda_1}{\Lambda} \  \hat{Q} \ \hat{L} \ \hat{d}^c \ \hat{\nu}^c 
\ + \  \frac{\lambda_2}{\Lambda} \  \hat{u}^c \ \hat{d}^c \ \hat{d}^c \ \hat{\nu}^c \ + \
\frac{\lambda_3}{\Lambda} \  \hat{L} \ \hat{L} \ \hat{e}^c \ \hat{\nu}^c. 
\end{equation}
As one expects, the R-parity violating terms in ${\cal W}_{BL}$, Eq.(2), are not allowed before symmetry breaking. 

Now, in the context of the minimal theory there is only one possibility to break the gauge symmetry, $G_{B-L}$, 
to the SM symmetry. We have to give a vacuum expectation value to the only fields which can be responsible 
for B-L symmetry breaking, the right-handed sneutrinos $\tilde{\nu}^c$. Therefore, one can say that R-parity 
must be spontaneously broken and one expects lepton number violating signatures at the LHC. Here the scale for 
B-L and R-parity violation is defined by the supersymmetric scale. It is easy to prove that we can give a vev 
to the fields $\tilde{\nu}^c$ in a consistent way, using the superpotential ${\cal W}_{B-L}$, the D-terms and the soft terms in
\begin{equation}
V_{soft} \supset  M_{\tilde{L}}^2  |\tilde{L}|^2  \ + \ M_{\tilde{\nu}^c}^2 |\tilde{\nu}^c|^2 \ + \  \left( A_\nu \ \tilde{L}  \ H_u \ \tilde{\nu}^c \ + \  \rm{h.c.} \right).
\end{equation} 
One finds that the VEV for the right-handed sneutrinos is given by
\begin{equation}
\left< \tilde{\nu}^c \right> = \sqrt{-  \frac{8 \ M_{\tilde{\nu}^c}^2}{g_{BL}^2}},
\end{equation} 
with $M_{\tilde{\nu}^c}^2 < 0$, and
\begin{equation}
\left< \tilde{\nu} \right> = \frac{\left< \tilde{\nu}^c \right>}{\sqrt{2}} \frac{ Y_\nu \ \mu \ v_d \ - \ A_\nu \ v_u}{\left( M_{\tilde{L}}^2 - \frac{g_{BL}^2}{8} \left< \tilde{\nu}^c \right>^2 \right)}.
\end{equation}
Here $g_{BL}$ is the B-L gauge coupling. The sfermion masses are different in this scenario because 
we have new contributions from the $B-L$ D-term and we can have 
tachyonic slepton masses if the $Z_{B-L}$ mass is large. Therefore, in order to have a realistic spectrum we have to 
impose the relation $M_{Z_{BL}} < \sqrt{2} \ M_{\tilde{L}}$~\cite{FileviezPerez:2012mj}.
The spectrum for neutrinos is peculiar in this theory. We have shown in Ref~\cite{Barger:2010iv} 
(see also Refs.~\cite{Mohapatra:1986aw,Ghosh:2010hy}) 
that the spectrum for neutrinos is different because there are five light neutrinos, the SM neutrinos and two sterile neutrinos. 
It has been pointed out in Ref.~\cite{FileviezPerez:2012mj} that the invisible decay of the $Z_{BL}$ is modified 
and maybe one can test this property at the LHC.

It is important to mention that after symmetry breaking one obtains the bilinear 
lepton number violating terms at the renormalizable level, while the baryon 
number violating interactions are suppressed by the cutoff scale. Then, 
assuming a large cutoff, $\Lambda \sim 10^{17}$ GeV, one can satisfy the 
bounds on the proton decay lifetime. 

The testability of this theory at the LHC has been investigated in Ref.~\cite{FileviezPerez:2012mj}, where we have discussed 
the most interesting signatures at the LHC in different LSP scenarios. Here we will mention the most striking signals 
which can help us to test the lepton number and spontaneous R-parity violation. We focus on the case where the lightest 
supersymmetric particle is the neutralino. In this case one can have the following signals:

$$pp \  \to \gamma^*, Z^*, Z_{BL}^* \ \to \ \tilde{e}^*_i \tilde{e}_i \  \to  \ e^{\pm}_i \ e^{\mp}_i \ e^{\mp}_j \ e^{\mp}_k \ 4j.$$
\begin{figure}[tb]
\includegraphics[scale=1,width=7.0cm]{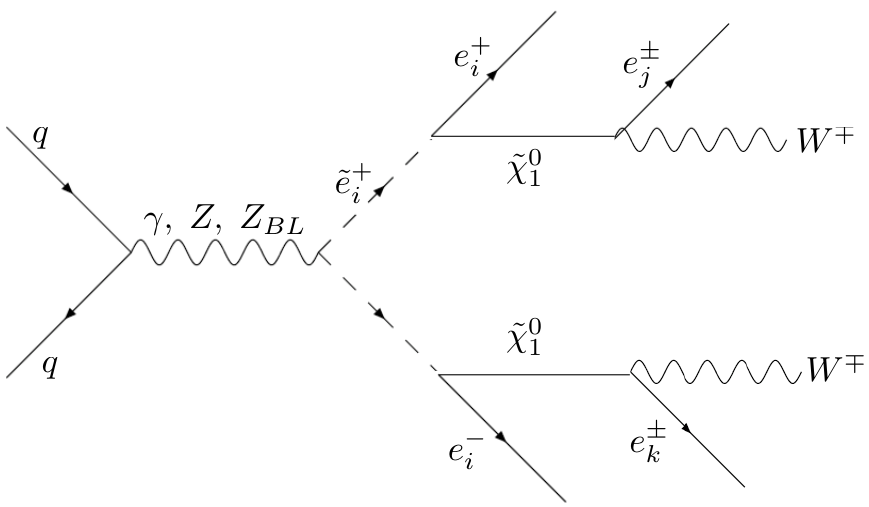}
\caption{Signals with four leptons and two W's. In order to observe lepton number violation we use the jets from the W decays.}
\label{figure1}
\end{figure}

These signals with four leptons, three of them with the same electric charge, and four jets are background free. 
See Fig. 1 for the topology of these events.
We have investigated in detail the neutralino decays taking into account the constraints coming from neutrino 
physics and the slepton decays, showing that with a luminosity of $10 \  \rm{fb}^{-1}$ we can have several events.
We have shown in Ref.~\cite{FileviezPerez:2012mj} that there are many solutions where the neutralino is long 
lived giving raise to displaced vertices. In Fig. 2 we can see that the production cross section can be above 
1 fb when the slepton mass is below 400 GeV. In order to understand the discovery reach at the LHC we show 
in Fig. 2 (right panel) the number of events for different values of the slepton mass and branching ratios. 
In this way we can see the possibility to have a few events in large region of the parameter space, 
for more details see Ref.~\cite{FileviezPerez:2012mj}.
\begin{figure}[tb]
\includegraphics[scale=1,width=6.0cm]{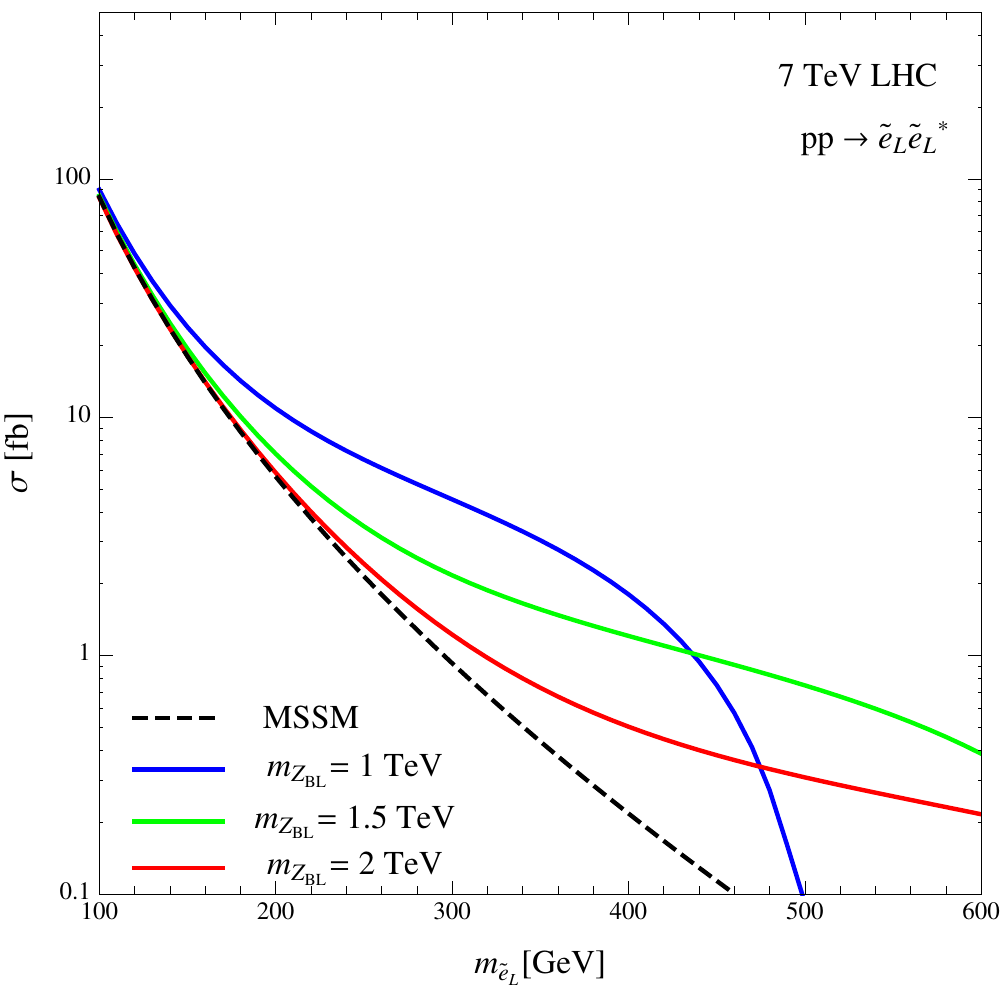}
\includegraphics[scale=1,width=6.65cm]{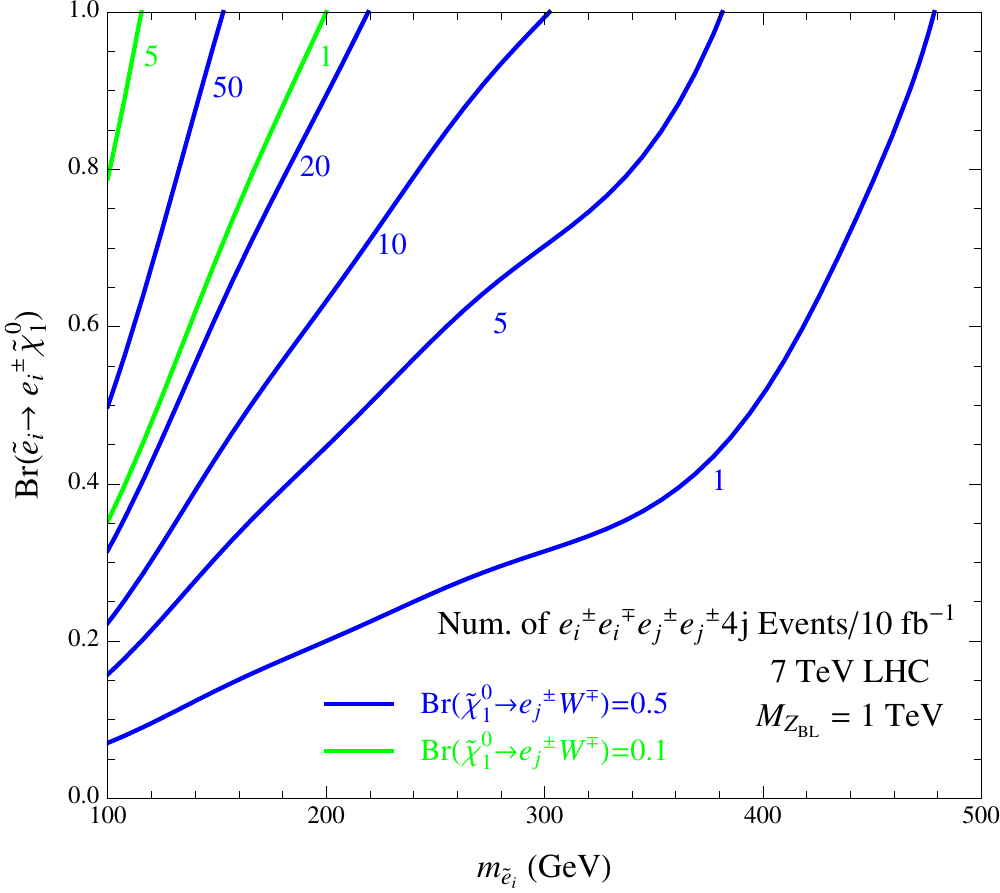}
\caption{Drell-Yan production cross sections for the charged sleptons. The dashed line corresponds to
the prediction in the MSSM and the solid lines show the results in our model for different values of the $Z_{BL}$ 
mass. In the right panel we show the number of events~\cite{FileviezPerez:2012mj}.}
\label{figure2}
\end{figure}

In summary, we can say that the minimal supersymmetric B-L theory~\cite{Barger:2008wn} predicts:
\begin{itemize}

\item R-parity must be spontaneously broken.

\item The B-L and R-parity violating scales are determined by the supersymmetry breaking scale.

\item The theory predicts two light sterile neutrinos.

\item Lepton number violating signals and displaced vertices at the LHC.

\end{itemize}
We would like to mention that these results are valid for the simplest theories where B-L is part of gauge 
symmetry and the idea was realized for the first time in Ref.~\cite{FileviezPerez:2008sx}.
It is important to emphasize again that in this case one needs to assume a large cutoff, i.e. a desert, 
in order to satisfy the constraints from proton decay. In the next sections we will discuss a 
model where one can break B and L at the low scale without generating any contribution to proton decay.
\subsection{II. R-parity Conservation at the LHC}
Now we discuss the minimal theory where one can explain dynamically the conservation of R-parity.
We have discussed in the previous section that in the simplest B-L theory one should break R-parity. 
Then, in order to have R-parity conservation we need to go beyond the minimal model and add extra 
Higgses to give mass to the new neutral gauge boson in the theory. The superpotentional of a simple 
theory for R-parity conservation is given by
\begin{equation}
{\cal W}_{RpC} = {\cal W}_{B-L} \ + \ \mu_X  \hat{X} \hat{\bar{X}} \ + \ f \  \hat{\nu}^c \hat{\nu}^c \hat{X}.
\end{equation}
We also could have two different scenarios: In the first case the new Higgses do not generate neutrino 
mass~\cite{Perez:2011zx}, or we can generate the mass for the $Z_{BL}$ gauge boson 
through the Stueckelberg mechanism~\cite{Feldman:2011ms}. Here we discuss the scenario 
where the neutrinos are Majorana fermions and have the implementation of the 
radiative symmetry mechanism to break the local B-L gauge symmetry.
\begin{figure}[h]
\includegraphics[scale=1,width=7.5cm]{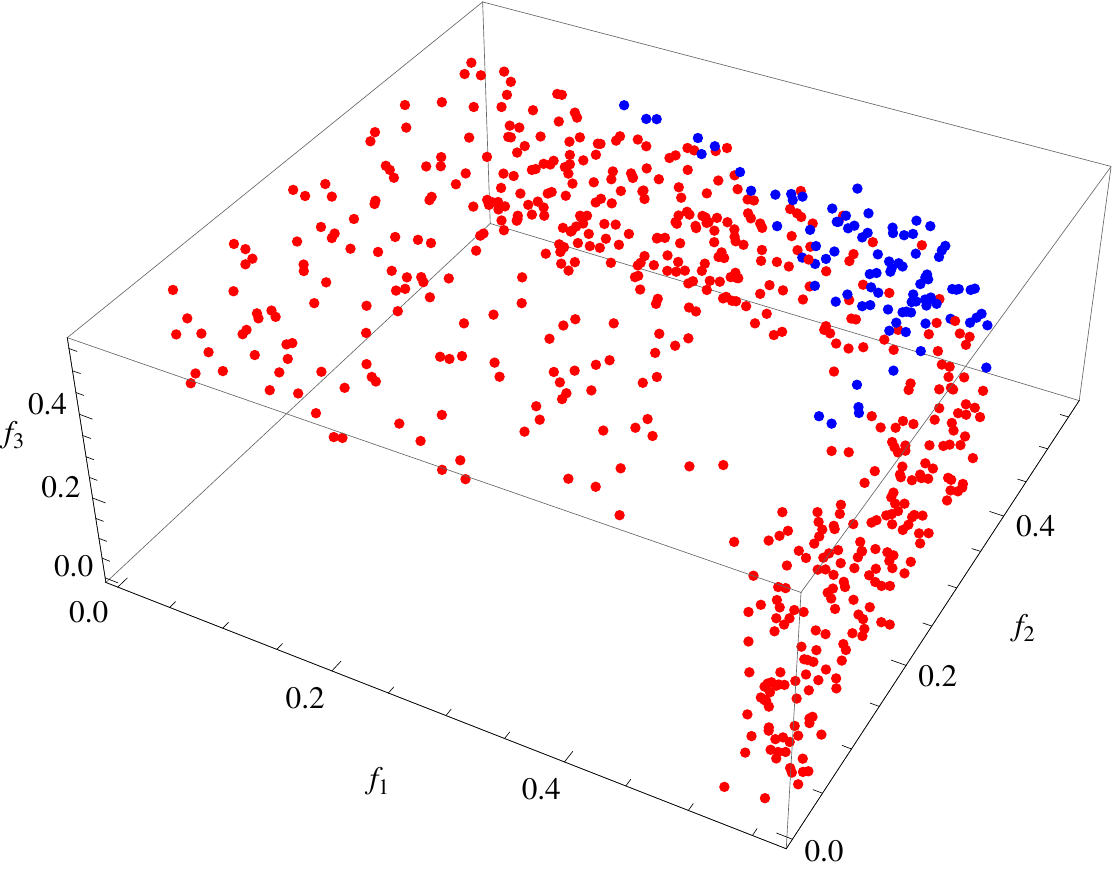}
\caption{The state of the $B-L$ breaking vacuum in the $f_1 - f_2 - f_3$ space with $M_0 = 2$ TeV, 
$M_{1/2} = 200$ GeV and $A_0 = 0$. Blue dots indicate R-parity conservation while red dots 
R-parity violation, the latter appears five times more often. The key point is that only fairly 
degenerate values of $f_i$ (and therefore the right-handed neutrino masses) allow for R-parity conservation~\cite{FileviezPerez:2011kd}.}
\label{figure3}
\end{figure}
\begin{figure}[h]
\includegraphics[scale=1,width=7.0cm]{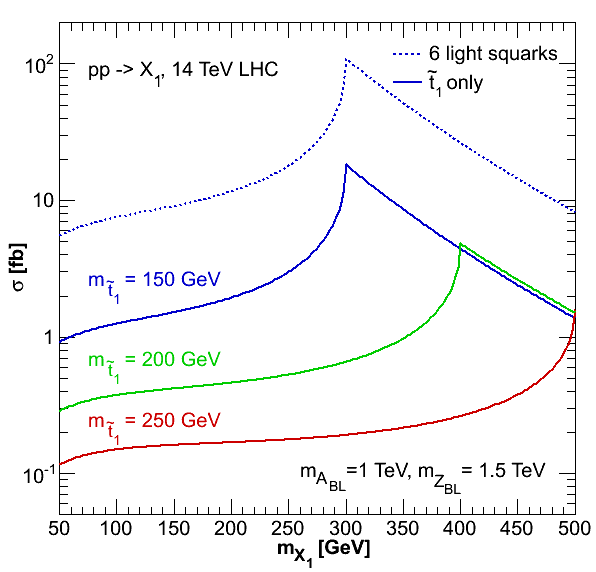}
\includegraphics[scale=1,width=6.7cm]{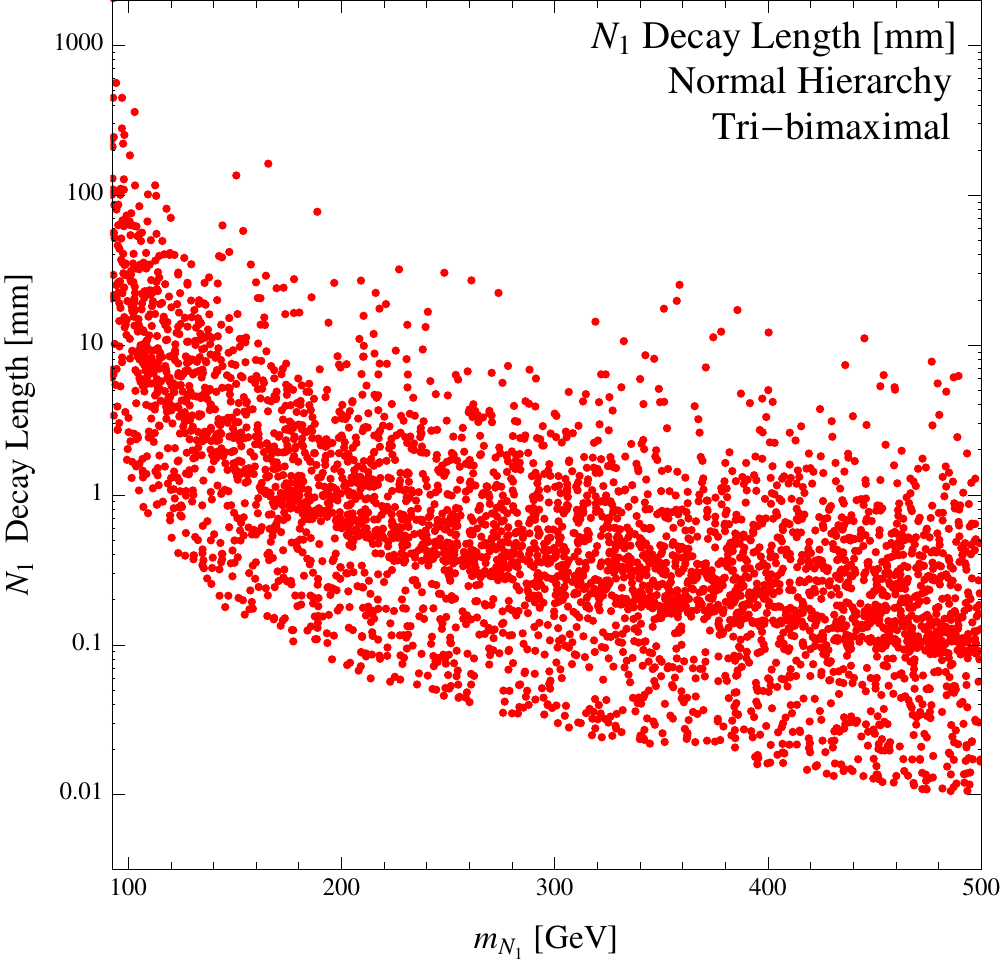}
\caption{Production cross section $pp \to X_1$ for 14 TeV assuming the input parameters listed in the plot. 
In the right panel we show right-handed neutrino decay length when the neutrino spectrum has a Normal 
Hierarchy~\cite{FileviezPerez:2011kd}.}
\label{figure4}
\end{figure}

In this theory we can have two different vacua which break B-L.
a) R-parity Conserving Vacua: In this case only the Higgses get a VEV, 
$\left< X \right> \neq 0$ and $\left< \bar{X} \right> \neq 0 $. 
b) R-parity Violating Vacua: The right-handed sneutrinos get 
a VEV. In Fig. 3 we show the possible solutions in the 3D $f_1 - f_2 - f_3$
space assuming boundary conditions~\cite{FileviezPerez:2010ek},
$A_0=0, \ M_0=2 \ \rm{TeV}, M_{1/2}=200 \ \rm{GeV}$. The blue and 
red points correspond to the solutions when R-parity is conserved and violated, respectively. 
Therefore, one can appreciate in Fig. 3 that if we stick to the radiative symmetry 
mechanism and universal boundary conditions R-parity is spontaneously broken 
in the majority of the parameter space. This is an interesting result which hints 
at the possibility that R-parity violation is present in this theory. 
We should mention that this simple theory for R-parity conservation predicts that the B-L symmetry 
breaking scale and the right-handed neutrino masses are determined by the SUSY scale, and 
one expects lepton number violating signals from the Higgs decays. 

This theory could be tested at the LHC if we understand the production mechanisms and decays of the new B-L Higgses, 
$X_1, X_2$ and $A_{BL}$. We have investigated the testability of this model at the LHC 
in Ref.~\cite{FileviezPerez:2011kd} and found that the signals with same-sign leptons 
and two jets  can be observed if we have high luminosity:
$$ pp \to X_1 \to N \ N \to WW \ e_i^{\pm} e_j^{\pm} \to e_i^{\pm} e_j^{\pm} \ 4j$$
In this case the number of events with two muons and four jets can be estimated as
\begin{eqnarray}
N_{2\mu 4j} &=& \sigma (pp \to X_1) \times \rm{Br} (X_1 \to N N) \times 2 \rm{Br} (N \to W \mu)^2  \times \rm{Br} (W \to 2j)^2 \times {\cal{L}}\\
& \approx & 5 \ \rm{fb} \times (1/3) \times 2 (1/4)^2 \times (6/9)^2 \times 100 \  \rm{fb}^{-1} \approx 9.
\end{eqnarray}
In Fig. 4 we show the cross section for the single Higgs production (left panel) and the decay length of the right-handed neutrinos 
(right panel). We can see that one has displaced vertices in a large fraction of the parameter space.
Therefore, we can have signals with two same-sign leptons and displaced vertices which are background free.
The Higgs pair production: $p p \  \to \ Z_{BL}^* \  \to \ X_1 \ A_{BL} \  \to \ N N N N$, is very important to test this 
model and its prediction is independent of the supersymmetric spectrum. See also Ref.~\cite{O'Leary:2011yq} for the 
study of the full spectrum assuming boundary conditions at the high scale. 

Now, we can say that it is very easy to write down a theory for R-parity conservation at the TeV scale 
and as in the previous model we can have interesting lepton number violating signals. 
In this context we can have the implementation of the radiative symmetry mechanism 
for the electroweak symmetry and the B-L gauge symmetry. The testability of this theory 
at the LHC could shed light on the connection between the cold dark matter 
in the universe and the possible observation of missing energy at the LHC.
\subsection{III. B and L as Local Gauge Symmetries}
In the previous sections we have discussed the simplest theory where one can understand the conservation or violation of $R$-parity. 
Unfortunately, in these theories we have to postulate the desert between the TeV scale and the grand unified scale in order to 
suppress the dimension five operators for proton decay. Now, we would like to discuss a simple theory where the local baryon 
and lepton numbers are local symmetries spontaneously broken at the TeV scale. This theory is based on the gauge symmetry~\cite{FileviezPerez:2011pt}
\begin{displaymath}
G_{BL}=SU(3)_C \bigotimes SU(2)_L \bigotimes U(1)_Y \bigotimes U(1)_{B} \bigotimes U(1)_L
\end{displaymath}
We refer to this model as the ``BLMSSM". In this context there are no dangerous operators mediating proton decay because 
the lepton number is broken in an even number while the baryon number violating operators can change B by one unit.
In this context the anomaly cancellation requires the presence of new families. In this context there is no flavour violation at tree level 
and in order to avoid Landau poles at the low scale we generate vector-like masses for the new quarks. We have investigated the predictions 
for the light Higgs boson mass showing that we can satisfy the experimental bounds without assuming a large stop mass and left-right 
mixing~\cite{FileviezPerez:2012iw}. Also we can modify the current LHC bounds on the supersymmetric spectrum due to the presence of the baryon number violating interactions.

In this model we have the chiral superfields of the MSSM, and in order to cancel the B and L anomalies we need a vector-like family:
$\hat{Q}_4$, $\hat{u}_4^c$ ,  $\hat{d}_4^c$, $\hat{L}_4$,  $\hat{e}_4^c$,  $\hat{\nu}^c_4$ and $\hat{Q}_5^c$, $\hat{u}_5$, $\hat{d}_5$,
$\hat{L}_5^c$, $\hat{e}_5$, $\hat{\nu}_5$. The superpotential of the model is given by
\begin{equation}
{\cal W}_{\rm{B}}^{\rm{L}}={\cal W}_{0} \ + \  {\cal W}_{\rm{B}} \ + \  {\cal W}_{\rm{L}} \ + \ {\cal W}_{X} \ + \ {\cal W}_{5},
\end{equation}
where
\begin{equation}
{\cal W}_{0}=Y_u \hat{Q} \hat{H}_u \hat{u}^c \ + \ Y_d \hat{Q} \hat{H}_d \hat{d}^c \ + \ Y_e \hat{L} \hat{H}_d \hat{e}^c \ + \ \mu \hat{H}_u \hat{H}_d,
\end{equation}
is the R-parity conserving MSSM superpotential and
\begin{eqnarray}
{\cal W}_{\rm{B}}&=&\lambda_{Q}  \hat{Q}_4  \hat{Q}_5^c \hat{S}_B \ + \  \lambda_{u}  \hat{u}_4^c  \hat{u}_5 \hat{\bar{S}}_B \ + \  \lambda_{d}  \hat{d}_4^c  \hat{d}_5 \hat{\bar{S}}_B \ + \   \mu_{B} \hat{\overline{S}}_B \hat{S}_B \nonumber \\
&+& Y_{u_4}  \hat{Q}_4 \hat{H}_u \hat{u}^c_4 \ + \ Y_{d_4} \hat{Q}_4 \hat{H}_d \hat{d}_4^c \ + \ Y_{u_5}  \hat{Q}^c_5 \hat{H}_d \hat{u}_5 \ + \ Y_{d_5} \hat{Q}_5^c \hat{H}_u \hat{d}_5.
\end{eqnarray}
The new quark superfields acquire TeV scale masses once the $S_B$ and $\bar S_B$ Higgs fields acquire a VEV. 
In the leptonic sector one has the following interactions
\begin{eqnarray}
{\cal W}_{\rm{{L}} }&=& Y_{e_{4}}  \  \hat{L}_4 \hat{H}_d \hat{e}^c_4 \ + \  Y_{e_{5}}  \  \hat{L}_5^c \hat{H}_u \hat{e}_5 \ + \ Y_{\nu_{4}} \  \hat{L}_4 \hat{H}_u \hat{\nu}^c_4
                         +  Y_{\nu_{5}}  \  \hat{L}_5^c \hat{H}_d \hat{\nu}_5 \nonumber \\
                          &+&   Y_{\nu}  \  \hat{L}  \hat{H}_u \hat{\nu}^c \ + \  \lambda_{\nu^c}  \  \hat{\nu}^c  \hat{\nu}^c  \hat{\overline{S}}_L \ + \  \mu_{L} \hat{\overline{S}}_L \hat{S}_L.
                          \label{Wlept}
\end{eqnarray}
Here we have an implementation of the seesaw mechanism for the light neutrino masses once the $\bar S_L$ field acquires a VEV, while the new neutrinos have Dirac mass terms.
In order to avoid the stability for the new quarks we add the fields, $\hat{X}$ and $\hat{\bar{X}}$, which have the following interactions
\begin{eqnarray}
{\cal W}_{\rm{X}}&=&\lambda_{1}  \hat{Q}  \hat{Q}_5^c \hat{X} \ + \  \lambda_{2}  \hat{u}^c  \hat{u}_5 \hat{\bar{X}} \ + \  \lambda_{3}  \hat{d}^c  \hat{d}_5 \hat{\bar{X}} \ + \  \mu_{X} \hat{\overline{X}} \hat{X},
\end{eqnarray}
where the baryon number for the new fields are: $B_X=2/3 + B_4=-B_{\bar{X}}$, and if we assume that they do not get a VEV the lightest one can be a dark matter candidate even if R-parity is violated, see Ref.~\cite{Arnold:2012fm} for details. For any value of the baryonic charges of the new fermions, which satisfy the anomaly conditions, the Higgses $\hat{S}_B$ and
$\hat{\overline{S}}_B$ have charges $1$ and $-1$, respectively. Then, one can write the following dimension five operator which gives rise to baryon number violation once the local baryonic symmetry is broken through the VEV of $S_B$:
\begin{eqnarray}
\label{bnv}
{\cal W}_5 &=& \frac{a_1}{\Lambda} \hat{u}^c \hat{d}^c \hat{d}^c \hat{S}_B.
\end{eqnarray}
Therefore, after breaking $U(1)_B$ we find the so-called $\lambda^{''}$ MSSM interactions which can modify the current LHC bounds on the supersymmetric mass spectrum. 

The B violating interactions can modify the signatures at the LHC. For example if the 
squark is the LSP it can be long-lived and form bounded states. Now, if we compute the decay length of a squark one finds
\begin{equation}
\rm{L} (\tilde{q}_i \to q_j q_k) \ > \  1 \ \rm{mm} \left(\frac{10^2 \  \rm{GeV}} {M_{\tilde q}} \right) \left(\frac{10^{-7}}{\lambda''}\right)^2.
\end{equation}
Therefore, the squark will form bounded states but it will decay inside the detector. In this case we have used the bound from 
cosmology~\cite{Arnold:2012fm}, and it is possible to have displaced vertices as well when the stop (sbottom) has mass around 100 GeV.  
For example, we can have signals with four jets from the decays of a stop or a sbottom
$$pp \ \to \ \tilde{t}^*  \tilde{t} \  \to \  4 j, \  pp \ \to \ \tilde{b}^*  \tilde{b} \  \to \  4 j.$$
Therefore, one can avoid the LHC constraints coming from the searches for multijets and missing energy~\cite{Arnold:2012fm}.

This theory also predicts light leptons which modify the predictions for the Higgs mass. 
In Fig. 5 we show the constraints on the stop mass and the left-right mixing in the stop sector 
in the MSSM and in the BLMSSM. Here we use $\tan \beta$ between 2 and 6 and see 
Ref.~\cite{Arnold:2012fm} for the other input parameters. Thanks to the existence 
of these new leptons one can satisfy the experimental bounds on the Higgs bounds 
even when the left-right mixing in the stop sector is small.
\begin{figure}[h]
\includegraphics[scale=1,width=8.4cm]{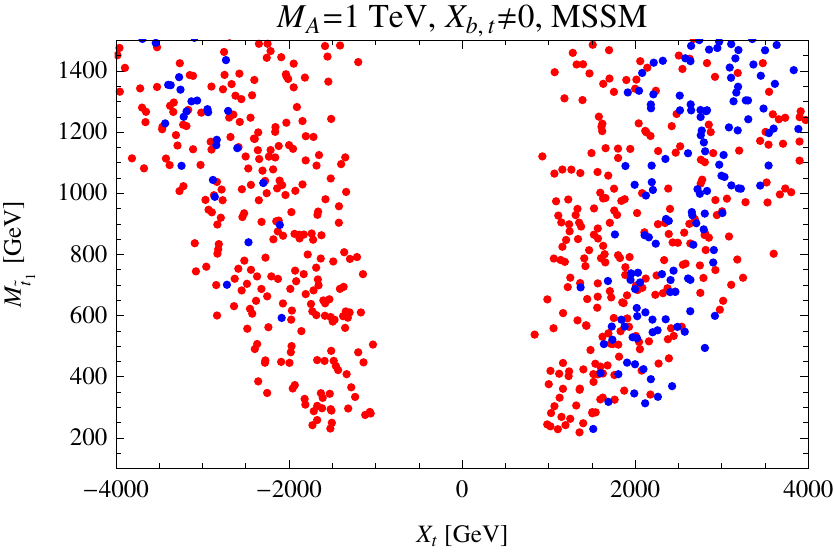}
\includegraphics[scale=1,width=8.4cm]{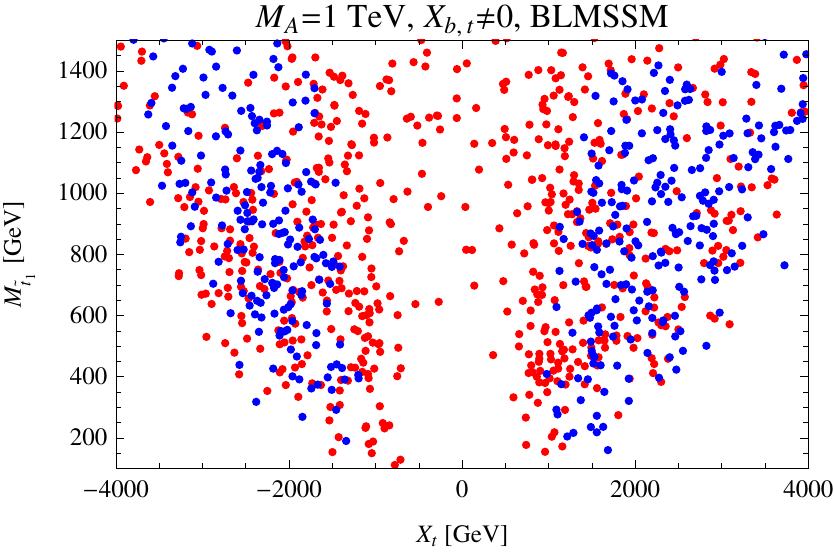}
\caption{Allowed parameter space in the MSSM and BLMSSM in the plane of lightest stop mass versus the left-right
mixing in the stop sector. We use as input parameters $M_{\nu_4} = M_{\nu_5} = 90$ GeV and 
$M_{e_4} = M_{e_5} = 100$ GeV. In the MSSM we compute the Higgs mass at two-loops and 
in the BLMSSM we have the extra one-loop contributions. The red points correspond to the range when the 
Higgs mass is between 115 GeV and 122 GeV, while the blue points correspond to the range, 
122 GeV - 128 GeV. We use $M_{\tilde g} = 1$ TeV as the gluino mass~\cite{Arnold:2012fm}.}
\label{figure5}
\end{figure}
The Higgs decays have been investigated in great details in Ref.~\cite{Arnold:2012fm}, 
where we have shown that the Higgs decay into two photons is suppressed and 
one could rule out this model in the near future if the recent Higgs signals, around $M_h \approx 125$ GeV, 
are confirmed by the LHC experiments.

In summary, we can say that it is possible to define a simple theory for the spontaneous breaking of B and L at the TeV 
scale in agreement with the experiments. The theory predicts baryon number at the low scale which 
could modify LHC bounds on sfermion masses. The lepton number is broken in an even number 
and one also expects lepton number violating signals at colliders. The new light leptons change 
the prediction for the Higgs mass and we could have a light supersymmetric spectrum.
\subsection{IV. 4D GUTs and R-parity Conservation}
In the context of a SUSY GUT in four dimensions based 
on the SU(5) gauge symmetry is not possible to understand the conservation 
of R-parity. In this context we can write the following interactions
\begin{equation}
{\cal W}_{SU(5)} \supset \epsilon_i \hat{\bar 5}_i 5_H \ + \ \lambda_{ijk} \hat{10}_i \hat{\bar 5}_j \hat{\bar 5}_k \ + \  \eta_i \hat{\bar 5}_i 24_H 5_H,
\end{equation} 
which break matter parity. Here $\hat{\bar 5}_i$ and $\hat{10}_i$ are the matter superfields, while $\hat{5}_H$, $\hat{\bar{5}}_H$ 
and $\hat{24}_H$ are the Higgs superfields.

Some of the $SO(10)$ scenarios provide a framework to understand the conservation of R-parity.
Unfortunately, one cannot use the $16_H$ Higgs to generate fermion masses and explain 
why the right-handed sneutrino does not get a vacuum expectation value. In the context 
of SO(10) we need to use large representations, $\hat{126}_H$, $\hat{\bar{126}}$, 
and $\hat{210}_H$ (or $\hat{54}_H$ and $\hat{45}_H$) to show that R-parity is conserved
and one can generate fermion masses at the renormalizable level. See 
Ref.~\cite{Aulakh:2000sn} for a discussion of this issue in SO(10). 
We can say that there is no a simple grand unified theory in four dimensions 
where we can understand the conservation of matter parity.
\section{Summary}
We have presented the simplest gauge theories for the conservation of R-parity in supersymmetry.
It has been shown that the minimal theory based on the B-L gauge symmetry predicts that R-parity must be 
spontaneously broken at the TeV scale. The B-L and R-parity violating scales 
are determined by the supersymmetry breaking scale and the theory predicts two light sterile neutrinos.
In this context we expect lepton number violating signals and displaced vertices at the LHC, 
the most striking signals are the channels with four leptons, three of them with the same electric charge, 
and four jets.

The minimal theory for R-parity conservation provides a framework to implement the radiative symmetry 
mechanism and dynamically generate neutrino masses. In this case, assuming universal boundary 
conditions for the soft terms, we have shown that R-parity is spontaneously broken in the majority of the parameter space. 
Here the B-L symmetry breaking scale and the right-handed neutrinos masses are determined by the SUSY scale. 
Also one expects lepton number violating signals from the Higgs decays and displaced vertices from the 
right-handed neutrino decays.

In order to avoid a desert between the TeV scale and the grand unified scale, but still satisfy the bounds 
on the proton decay lifetime, we have defined an interesting theory where the local 
baryon and lepton numbers are spontaneously broken at the supersymmetric scale. 
This theory predicts baryon number violation at the low scale which can modify the LHC bounds 
on the supersymmetric spectrum. The new light leptons in the theory increase the Higgs mass 
without assuming very heavy stops or a large left-right mixing in the stop sector.
The possibility to understand the conservation of R-parity in grand unified theories was mentioned.
\begin{theacknowledgments} 
I would like to thank the organizers of the GUT2012 Workshop at the Yukawa Institute for Theoretical Physics 
in Kyoto for the invitation and warm hospitality. It is a pleasure to thank my collaborators 
J. M. Arnold, V. Barger, D. Feldman, B. Fornal, P. Nath, S. Spinner, M. K. Trenkel and M. B. Wise for many discussions 
and enjoyable collaborations. This work has been supported by the James Arthur Fellowship, CCPP, New York University.
\end{theacknowledgments}

\bibliographystyle{aipproc}

\end{document}